\def\aa{{A\&A}}
\def\aas{{A\&AS}}
\def\aj{{AJ}}
\def\annrev{{ARA\&A}}
\def\apj{{ApJ}}
\def\apjs{{ApJS}}
\def\baas{{BAAS}}
\def\mnras{{MNRAS}}
\def\nat{{Nature}}
\def\pasj{{Proc.\ Astr.\ Soc.\ Japan}}
\def\pasp{{PASP}}

\newcommand	\cm	{\,{\rm cm}}
\newcommand	\erg	{\,{\rm ergs}}
\newcommand	\eV	{\,{\rm eV}}
\newcommand	\K	{\,{\rm K}}
\newcommand	\kms	{\,{\rm km\,s}^{-1}}
\newcommand	\lambdaref	{{\lambda_0}}
\newcommand	\micron	{\,\mu{\rm m}}
\newcommand     \gtsim  {\lower.5ex\hbox{$\; \buildrel > \over \sim \;$}}
\newcommand     \ltsim  {\lower.5ex\hbox{$\; \buildrel < \over \sim \;$}}
\newcommand	\yr	{\,{\rm yr}}
\newcommand	\s	{\,{\rm s}}
\newcommand	\sol	{{\odot}}

\documentclass[cup5b]{caps}
\usepackage{graphicx}
\usepackage{amssymb}
\usepackage{ociwsymp4}   %Use this for invited papers.

\HeadText{B. T. Draine}  %Enter author name; if three authors, list all three;
\begin{document}

\pagenumbering{arabic}

%Author names should be in captital letters
\author[]{B. T. DRAINE\\Princeton University Observatory}

\chapter{Interstellar Dust}

\begin{abstract}

In the interstellar medium of the Milky Way, 
certain elements -- e.g., Mg, Si, Al, Ca, Ti, Fe --
reside predominantly in interstellar dust grains.
These grains absorb, scatter, and emit electromagnetic radiation,
heat the interstellar medium by photoelectric emission, 
play a role in the ionization balance of the gas,
and catalyze the formation of molecules, particularly H$_2$.
I review the state of our knowledge of the composition and sizes of 
interstellar grains, including what we can learn from spectral features,
luminescence, 
scattering,
infrared emission,
and observed gas-phase depletions.
The total grain volume in
dust models which reproduce interstellar extinction
is significantly greater than estimated from observed depletions.

Dust grains might reduce the gas-phase D/H
ratio, providing an alternative mechanism to explain
observed variations in the gas-phase D/H ratio in the local interstellar
medium.  
Transport in dust grains
could cause elemental abundances
in newly-formed stars to differ
from interstellar abundances.

\end{abstract}

\section{Introduction}

Certain elements -- Si and Fe being prime examples --
are often extremely {\it depleted}
from the interstellar gas (Jenkins 2003).
The explanation for this
depletion is that the atoms that are ``missing'' from the gas phase
are located in
solid particles -- interstellar dust grains.  
The existence of this solid phase 
complicates efforts to determine elemental abundances, due to
the difficult-to-determine abundance in grains.

The concentration of certain elements in dust grains
also creates the possibility of selective transport of those elements 
through the gas, which could, at least in principle, lead to variations in
elemental abundances from point to point in the interstellar medium,
and possibly even differences between stellar and interstellar abundances.
Dust grains might even deplete deuterium from the gas, providing an
alternative explanation for observed variations in D/H ratios.

Observations of interstellar dust were recently reviewed by
Whittet (2003) and Draine (2003a).
The astrophysics of interstellar dust is a broad subject, encompassing
dust grain formation and destruction,
charging of dust grains, heating of interstellar gas by photoelectrons from
dust, electron transfer from dust grains to metal ions, the optics of
interstellar dust, optical luminescence and infrared emission from dust,
chemistry on dust grains, and the
dynamics of dust grains, including radiation-driven drift of grains relative
to gas, and alignment of dust by the magnetic field.
Introductions to the astrophysics of dust can be found in
Kr\"ugel (2003) and Draine (2004).

\section{Presolar Grains in Meteorites are Nonrepresentative}

Genuine interstellar grains are found in meteorites
(Clayton \& Nittler 2003), but, because
present search techniques rely on isotopic anomalies,
the grains that are found are limited to those formed in stellar winds
or ejecta.
Theoretical studies of grain destruction by supernova-driven blastwaves
lead to estimated grain lifetimes of only $\sim2-3\times10^8\yr$
(Draine \& Salpeter 1979; Jones et al. 1994); this, together with the
large depletions typically seen for elements like Si implies, that
the bulk of interstellar grain material must be regrown in the interstellar
medium (Draine \& Salpeter 1979, Draine 1990).

Because the 
typical interstellar grains don't bear a distinctive isotopic signature, 
we don't know how to identify them in meteorites.
We are therefore forced to study interstellar grains remotely.

\section{Interstellar Reddening}

Our knowledge of interstellar dust is largely derived from
the interaction of dust particles with electromagnetic
radiation: attenuation of starlight,
scattering of light, and emission of infrared
and far-infrared radiation.

The wavelength dependence of interstellar extinction 
tells us about both the size and composition
of the grains.  The extinction is best determined using the
``pair method'' -- comparison of the fluxes from two stars with
nearly-identical spectroscopic features (and therefore photospheric
temperature and gravity) but with one of the stars nearly unaffected
by dust.  With the assumption that the extinction goes to zero as
wavelength $\lambda\rightarrow\infty$, it is possible to determine
the extinction $A_\lambda$ as a function of wavelength
(see, e.g., Fitzpatrick \& Massa 1990, and references therein).

The extinction $A_\lambda$ is obviously proportional to the amount of dust,
but $A_\lambda/A_{\lambdaref}$,
the extinction normalized to some reference wavelength $\lambdaref$,
characterizes the {\it kind} of dust present, and its size distribution.
The quantity
$R_V\equiv A_V/(A_B-A_V)$ characterizes the slope of the extinction curve
between $V=0.55\micron$ and $B=0.44\micron$; small values of $R_V$ correspond
to steep extinction curves.

\begin{figure}
\centering
\includegraphics[width=8cm,angle=270]
{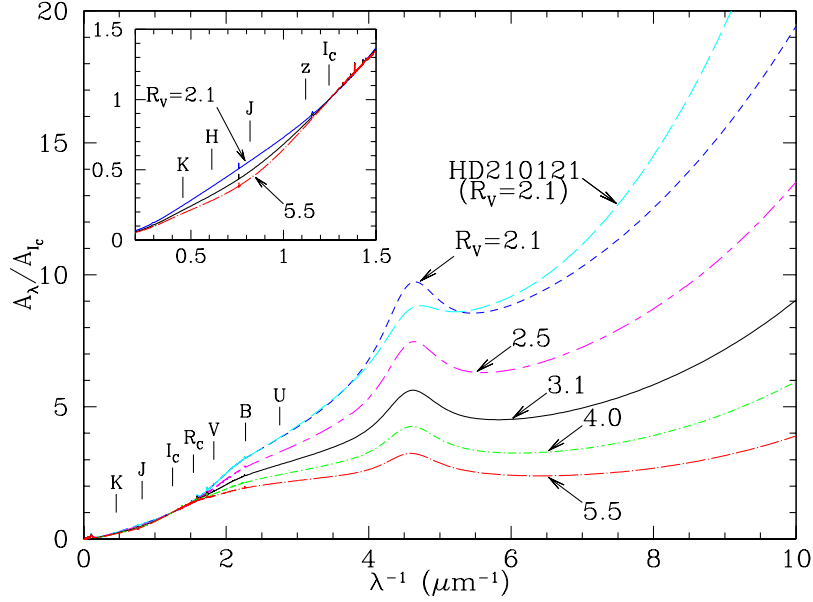}
%{f1.cps}
\caption{Extinction normalized to Cousins I band extinction for
$R_V$ values ranging from 2.1 to 5.5, using the
Fitzpatrick (1999) parameterization, plus diffuse interstellar bands
following Jenniskens \& Desert (1994).  
Also shown is an improved fit to
the extinction curve toward HD21021, providing one example of how
a sightline can deviate from the average behavior for the same value
of $R_V$.}
\label{fig:Alambda_over_AI}
\end{figure}

In principle, the function $A_\lambda/A_\lambdaref$ is unique to every
sightline,
but Cardelli et al.\ (1989) found that the
observed $A_\lambda/A_\lambdaref$ can be approximated by
a one-parameter family of curves:
$A_\lambda/A_{\lambdaref} = f(\lambda,R_V)$,
where they chose $R_V\equiv A_V/(A_B-A_V)$ as the parameter because it
varies significantly from one curve to another.
Cardelli et al.\ obtained functional forms for $f(\lambda,R_V)$ which
provided a good fit to observational data;
Fitzpatrick (1999) revisited this question and, explicitly correcting
for the
finite width of photometric bands, obtained a slightly revised set of
fitting functions.
Figure \ref{fig:Alambda_over_AI} shows the Fitzpatrick (1999) fitting
functions, using $I_C=0.802\micron$, the central wavelength of the
Cousins I band, as the reference wavelength.
The parameterization is shown for values of $R_V$ ranging from 2.1 to 5.5 ,
which spans the range of $R_V$ values encountered on sightlines through
diffuse clouds in the Milky Way.
Also shown is an empirical fit to the extinction measured toward
HD210121, showing how an individual sightline can deviate from the
one-parameter fitting function $f(\lambda,R_V)$.

Dust on sightlines with different values of $R_V$ obviously must have 
either different compositions or different size distributions, or both.
Also of interest is the total amount of dust per unit H.
This requires measurement of the total H column density
$N_{\rm H}\equiv N({\rm H})+2N({\rm H}_2)+N({\rm H}^+)$.
On most sightlines the ionized hydrogen is a small correction;
$N({\rm H})$ and $N({\rm H}_2)$ can be measured using 
ultraviolet absorption lines.

\begin{figure}
\centering
\includegraphics[width=8cm,angle=270]
{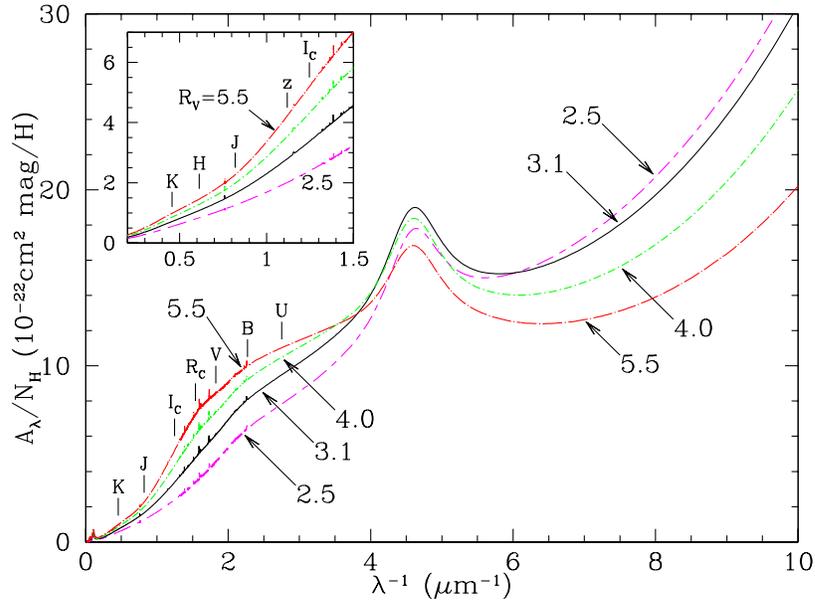}
%{f2.eps}
\caption{Extinction per unit H column density, for different $R_V$.
	From Draine (2003a)}
\label{fig:Alambda_over_NH}
\end{figure}

Rachford et al.\ (2002) determined $N_{\rm H}$ to
an estimated accuracy of better than a factor 1.5 on 14 sightlines.
It appears that $A_I/N_{\rm H}$ is positively correlated with $R_V$,
with 
\begin{equation}
\frac{A_I}{N_{\rm H}}\approx\left[2.96-3.55\left(\frac{3.1}{R_V}-1\right)\right]
\times10^{-22}{\rm cm}^2
\label{eq:A_I/N_H}
\end{equation}
providing an empirical fit (Draine 2003a). 
We can
use the Fitzpatrick (1999) parameterization 
and equation (\ref{eq:A_I/N_H}) to estimate $A_\lambda/N_{\rm H}$
for sightlines with different $R_V$.
The results are shown in 
Figure \ref{fig:Alambda_over_NH} -- sightlines with larger $R_V$ values
appear to have larger values of $A_\lambda/N_{\rm H}$ for
$\lambda^{-1} \ltsim 3\micron^{-1}$, and decreased values
for $\lambda^{-1} \gtsim 4\micron^{-1}$.
This is interpreted as resulting from coagulation of a fraction of
the smallest grains onto the larger grains;
loss of small grains decreases the ultraviolet extinction,
while adding mass to the larger grains
increases the scattering at $\lambda\gtsim0.3\micron$.

\newpage

\section{Spectroscopy of Dust in Extinction and Emission}

\subsection{PAH Emission Features \label{sec:PAHs}}

\begin{figure}
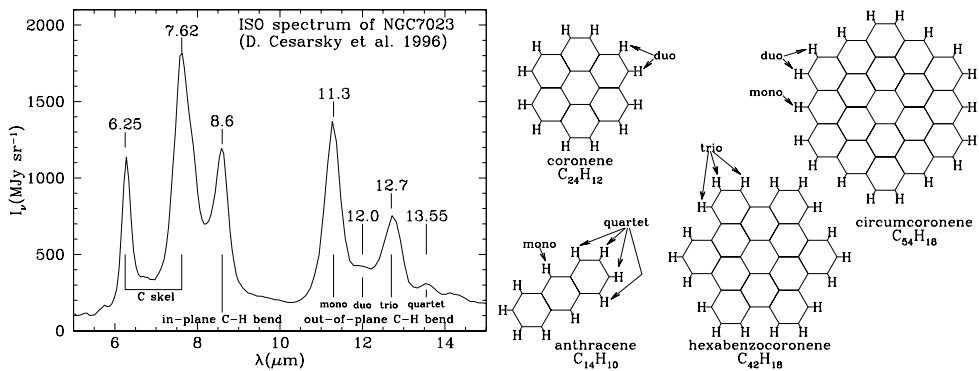

\centering
\hspace*{-1.5cm}
\includegraphics[width=4.8cm,angle=270]
{f3a.cps}
%{f3a.eps}
\includegraphics[width=4.8cm,angle=270]
{f3b.cps}
%{f3b.eps}
\caption{PAH emission features in the 5-15$\mu$m emission spectrum of
	the reflection nebula NGC~7023 (Cesarsky et al.\ 1996),
	and four PAH molecules, with examples of mono, duo, trio, and quartet
	H sites indicated.}
\label{fig:PAH}
\end{figure}

Interstellar dust glows in the infrared, with a significant fraction of
the power radiated at $\lambda\ltsim 25\micron$.  In reflection nebulae
the surface brightness is often high enough to permit spectroscopy of
the emission; the 5--15$\micron$ spectrum of NGC 7023 is shown in
Figure \ref{fig:PAH}.
The spectrum has 5 very conspicuous emission peaks, at
$\lambda=12.7, 11.3, 8.6, 7.6,$ and $6.25\micron$;
there is an additional emission peak at $3.3\micron$ (not shown here),
as well as weaker peaks at 12.0 and 13.6$\micron$.

These emission features are in striking agreement with the wavelengths of
the major optically-active vibrational modes for polycyclic aromatic
hydrocarbon (PAH) molecules: the 5--15$\micron$ features are labelled in
Figure \ref{fig:PAH}; the 3.3$\micron$ feature (not shown) is
the C-H stretching mode.
The vibrational excitation results from single-photon heating
(see \S\ref{sec:IR+FIR_emission}).
The strength of the observed emission requires that
PAH molecules be a major component of
interstellar dust.
Modeling the observed emission in reflection nebulae
indicates that the PAH species containing $\ltsim10^3$ C atoms contain
$\sim$40~ppm C/H -- approximately 15\% of $({\rm C/H})_\sol= 246\pm23$~ppm
(Allende Prieto et al 2002).

\subsection{2175\AA\ Feature: Graphitic C}

By far the strongest feature in the extinction curve is the 
2175\AA\ ``bump'' 
(see Figures \ref{fig:Alambda_over_AI} and \ref{fig:Alambda_over_NH}).
Stecher \& Donn (1965) pointed out that the observed feature coincided
closely with the position and width of absorption expected from 
small spheres of graphite.
Graphite consists of parallel sheets of graphene -- two-dimensional
hexagonal carbon lattices; adjacent graphene layers interact only
through a weak van der Waals interaction.

The 2175\AA\ feature arises from a $\sigma\rightarrow\sigma^*$ excitation
of a $\sigma$ orbital in graphene; in small
graphite spheres the feature strength corresponds to an oscillator
strength per C of 0.16 (Draine 1989), 
so that the observed 2175\AA\ feature requires
C/H$\approx$ 60~ppm 
$\approx0.25({\rm C/H})_\sol$
to account for the observed strength of
the 2175\AA\ feature.

In a large PAH molecule, the interior C atoms
have electronic orbitals closely resembling those in graphene, and
one therefore expects that large PAH molecules will have a
strong absorption feature peaking near 2175\AA,
with an oscillator strength per C expected to be close to the
value for graphite. 
It is of course interesting to note that the C/H in 
PAHs required to account for the 3--15$\micron$ IR emission is
$\sim2/3$ of that required to account for the 2175\AA\ bump.
Since it would be entirely natural to have additional PAHs containing
$> 10^3$ C atoms 
it is plausible that the 2175\AA\
feature could be entirely due to PAHs, in which case the PAHs contain
C/H $\approx$ 60~ppm.

The 2175\AA\ feature is suppressed in graphite grains with 
$a\gtsim0.02\micron$
because the grain becomes opaque throughout the 1800--2600\AA\ range.
There can therefore be additional ``aromatic'' C
within $a\gtsim0.02\micron$ grains.

\subsection{10 and 18$\mu$m Silicate Features}

Interstellar dust has a strong absorption feature at 9.7$\micron$.
While the precise composition and structure of the carrier remains
uncertain, there is little doubt that the 9.7$\micron$ feature is 
produced by
the Si--O stretching mode in silicates.
In the laboratory, 
crystalline silicates have multiple narrow features in
their $10\micron$ spectra, while amorphous silicate material
has a broad profile.

The interstellar 9.7$\micron$ 
feature is seen in absorption on a number of sightlines.
The observed profiles are broad and relatively featureless, 
indicative of amorphous silicate material.
The observed strength of teh absorption feature requires that much,
perhaps most, of interstellar Si atoms reside in silicates.
Li \& Draine (2001a) estimated that at most 5\% of interstellar
silicate material was crystalline.
Conceivably, a mixture
of a large number of different crystalline minerals (Bowey \& Adamson 2002)
could blend together to produce the observed smooth profiles,
although it seems unlikely that nature would have produced a blend of
crystalline types with no fine structure evident in either absorption or
emission, including emission in the far-infrared (Draine 2003a).
The $9.7\micron$ extinction profile does not appear to be ``universal'' --
sightlines through the diffuse ISM show a {\it narrower} feature than
sightlines through dense clouds (Roche \& Aitken 1984; Bowey et al.\ 1998).
Evidently the silicate material is altered in interstellar
space.

Identification of the 9.7$\micron$ feature as the Si--O stretching
mode is confirmed by the presence of a
broad feature centered at $\sim$18$\micron$ 
(McCarthy et al.\ 1980, Smith et al.\ 2000) that is
interpreted as the O--Si--O bending mode in silicates.

Spectral features of crystalline silicates are seen in some
circumstellar disks (Artymowicz 2000, Waelkens et al.\ 2000) and
some comets (Hanner 1999), but even in these objects only a minority
of the silicate material is crystalline (e.g., Bouwman et al.\ 2001).

\subsection{3.4$\mu$m C-H Stretch: Aliphatic Hydrocarbons}

Sightlines with sufficient obscuration reveal a broad
absorption feature at 3.4$\micron$ that is identified as the C--H
stretching mode in aliphatic (i.e., chain-like) hydrocarbons.
Unfortunately, it has not proved possible to identify the specific
aliphatic hydrocarbon material, and
the band strength of this mode varies significantly
from one aliphatic material to another.
Sandford et al.\ (1991) suggest that the 3.4$\micron$ feature is due to
short saturated aliphatic chains incorporating C/H $\approx$ 11~ppm,
while Duley et al.\ (1998) attribute the feature
to hydrogenated amorphous carbon (HAC) material containing C/H $\approx$
85~ppm.

\subsection{Diffuse Interstellar Bands}

In addition to narrow absorption features identified as atoms, ions,
and small molecules, the observed extinction
includes a large number of broader features -- known as the ``diffuse
interstellar bands'', or DIBs.
The first DIB was recognized over 80 years ago (Heger 1922),
and shown to be interstellar 70 years ago
(Merrill 1934).
A recent survey by Jenniskens \& Desert (1994) lists
over 154 ``certain'' DIBs, with another
52 ``probable'' features.
Amazingly, not a single one has yet been positively identified!

High resolution spectroscopy of the 5797\AA\ and 6614\AA\ DIBs
reveals fine structure that is consistent with rotational bands
in a molecule with tens of atoms (Kerr et al.\ 1996, 1998), and
there is tantalizing evidence that DIBs at 9577\AA\ and 9632\AA\
may be due to C$_{60}^+$ (Foing \& Ehrenfreund 1994; 
Galazutdinov et al 2000, but see also Jenniskens et al.\ 1997 and
Moutou et al.\ 1999)

It appears likely that some or all of the DIBs
are due to absorption in large molecules or ultrasmall grains.
As noted above, a large population of PAHs is required to account for
the observed IR emission, and it is reasonable to suppose that these
PAHs may be responsible for many of the DIBs.  What is needed now
is laboratory gas-phase absorption spectra
for comparison with observed DIBs.  Until we have precise wavelengths
(and band profiles) from gas-phase measurements, secure identification
of DIBs will remain problematic.

\subsection{Extended Red Emission}

Interstellar dust grains luminesce in the far-red, 
a phenomenon referred to
as the ``extended red emission'' (ERE).
The highest signal-to-noise observations are in reflection nebulae, where
a broad featureles emission band is observed to peak at wavelength
$6100\ltsim$ $\lambda_p \ltsim8200$\AA, with a FWHM in the range 600-1000\AA\
(Witt \& Schild 1985, Witt \& Boroson 1990).
The ERE has also been seen in H~II regions (Darbon et al.\ 2000), 
planetary nebulae (Furton \& Witt 1990), and
the diffuse interstellar medium of our Galaxy
(Gordon et al.\ 1998, Szomoru \& Guhathakurta 1998).

The ERE is photoluminescence: 
absorption of a starlight photon raises the grain to
an excited state from which it decays by spontaneous emission of
a lower energy photon.  
The reported detection of ERE from the diffuse interstellar
medium (Gordon et al.\ 1998, Szomoru \& Guhathakurta 1998)
appears to require that the interstellar dust mixture have
an overall photoconversion 
efficiency of order $\sim$10\% 
for photons shortward of $\sim$5000\AA.
If the overall efficiency is $\sim$10\%, the ERE carrier itself must contribute
a significant fraction of the overall absorption by interstellar dust at
$\lambda\ltsim5000$\AA.
  
Candidate ERE carriers which have been proposed
include PAHs (d'Hendecourt et al.\ 1986)
and silicon nanoparticles (Ledoux et al.\ 1998, Witt et al.\ 1998,
Smith \& Witt 2002).
While some PAHs are known to luminesce, 
attribution of the ERE to PAHs is difficult
because of nondetection of PAH emission from some regions where ERE is seen
(Sivan \& Perrin 1993, Darbon et al 2000) 
and nondetection of ERE in some reflection
nebulae with PAH emission (Darbon et al 1999).
Oxide-coated silicon nanoparticles
appear to be ruled out by nondetection of infrared emission at $\sim20\micron$
(Li \& Draine 2002a).  The search to identify the ERE carrier continues.

\subsection{X-Ray Absorption Edges}

\begin{figure}
\centering
\includegraphics[width=8cm,angle=0]
{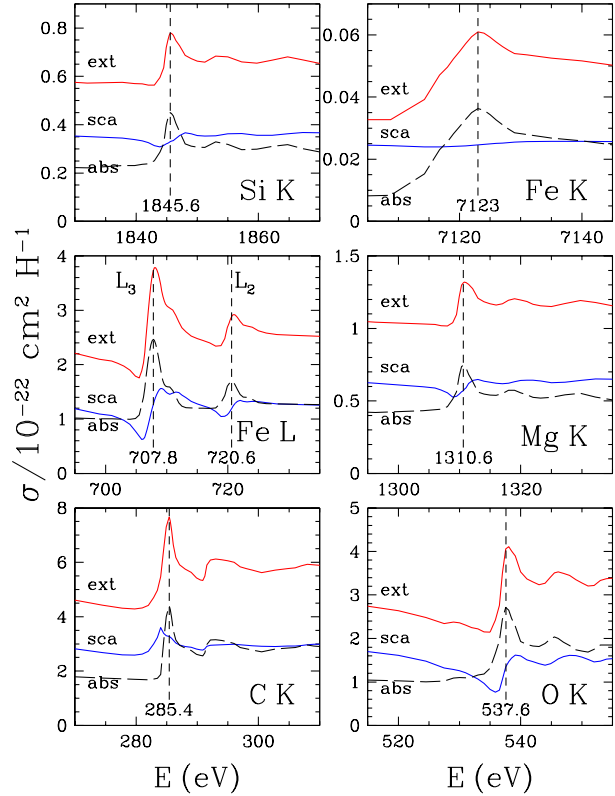}
%{f4.eps}
\caption{Scattering and absorption cross sections per H nucleon
	near principal X-ray absorption edges,
	for an interstellar dust model consisting of graphite and
	amorphous silicate grains (from Draine 2003c).}
\label{fig:NEXAFS}
\end{figure}
Dust grains become nearly transparent at X-ray energies, so that the
measured X-ray 
absorption is sensitive to {\it all} of the atoms, not just those
in the gas phase.  Consider, for example, K shell absorption by an oxygen
atom, where photoabsorption excites one of the 1s
electrons to a higher (initially vacant) energy level.
Transitions to a bound state ($2p$, $3p$, $4p$, ...) produce a series of
absorption lines, and transitions to unbound ``free'' 
(i.e, ``photoelectron'') states
result in a continuum beginning at the ``absorption edge''.
If the atom is in a solid, the absorption spectrum is modified
because the available bound states and free electron states are
modified by the presence of other nearby atoms.
High resolution X-ray absorption spectroscopy of interstellar matter
could thereby identify the chemical form in which elements
are bound in dust grains (Forrey et al.\ 1998, and references therein).

Figure \ref{fig:NEXAFS} shows the structure expected for scattering,
absorption, and extinction near the major X-ray absorption edges
in a grain model composed of $sp^2$-bonded carbon grains plus
amorphous MgFeSiO$_4$ grains (Draine 2003c).  Spectroscopy
with $\ltsim$1~eV energy resolution near these absorption
edges for both
interstellar sightlines and laboratory samples can
test this model for the composition of interstellar dust.
The Chandra X-Ray Observatory has measured the wavelength-dependent
extinction near the absorption edges of
O (Paerels et al.\ 2001, Takei et al.\ 2002) and O, Mg, Si, and Fe
(Schulz et al.\ 2002), but it has not yet proved possible to identify
the chemical form in which the solid-phase Mg, Si, Fe, and O reside.

\section{Infrared Emission Spectrum of Interstellar Dust
	\label{sec:IR+FIR_emission}
	}

\begin{figure}
\centering
\includegraphics[width=8cm,angle=270]
{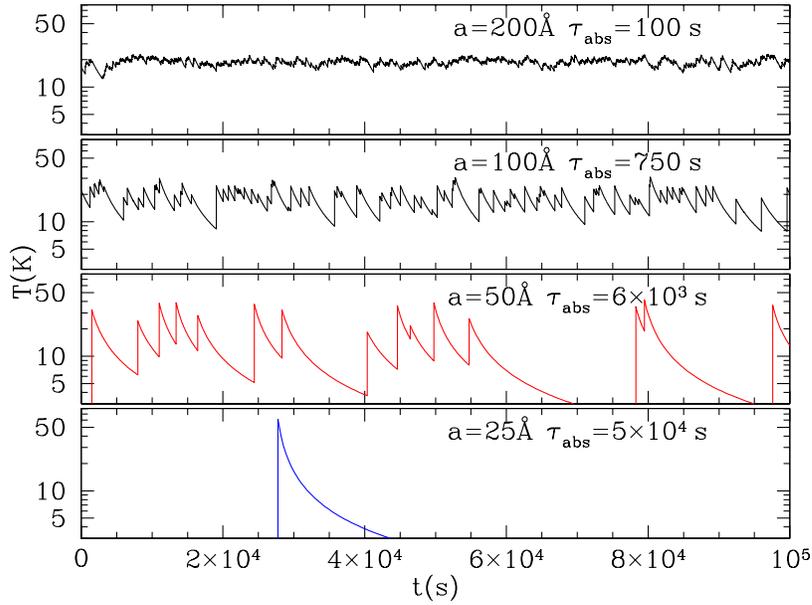}
%{f5.eps}
\caption{A day in the life of 4 carbonaceous grains, exposed to
	the average starlight background.
	$\tau_{\rm abs}$ is the mean time between photon absorptions.}
	\label{fig:Thist}
\end{figure}

Interstellar dust grains are heated primarily by absorption of starlight
photons.
A small fraction of the absorbed starlight energy goes into luminescence
or ejection of a photoelectron, but the major part of the absorbed starlight
energy goes into heating (i.e., vibrationally exciting) 
the interstellar grain material.  Figure \ref{fig:Thist} shows
grain temperature vs. time simulated for four sizes of carbonaceous grains
exposed to the average interstellar radiation field.
For grain radii $a\gtsim100$\AA, individual photon absorptions are
relatively frequent, and the grain heat capacity is large enough that
the temperature excursions following individual photon absorptions
are relatively small; it is reasonable to approximate the grain temperature
as approximately constant in time.
For $a\ltsim50$\AA\ grains, however, the grain is able to cool appreciably
in the time between photon absorptions; as a result, individual photon
absorption events raise the grain temperature to well above the mean value.
To calculate the time-averaged infrared emission spectrum for these grains,
one requires the temperature distribution function 
(see, e.g., Draine \& Li 2001).

\begin{figure}
\centering
\includegraphics[width=8cm,angle=270]
{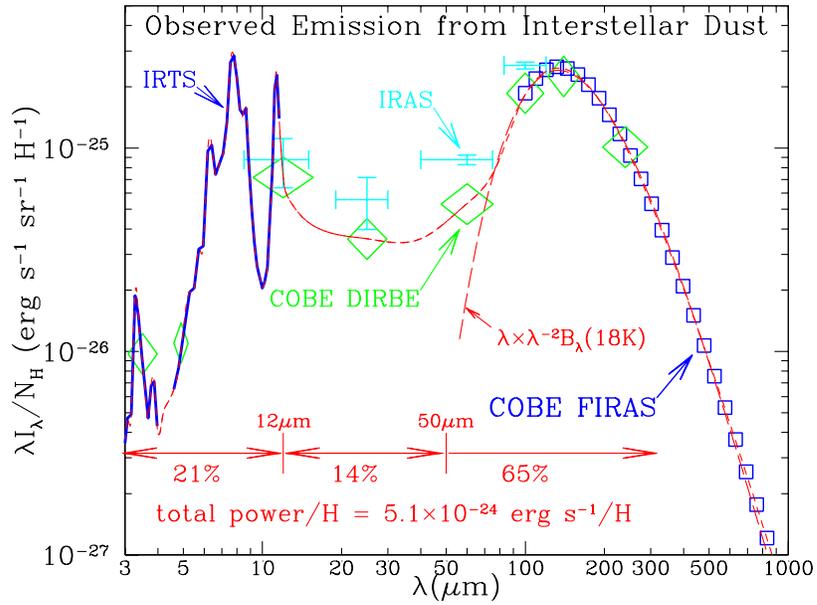}
%{f6.eps}
\caption{Observed emission spectrum of diffuse interstellar dust 
in the Milky Way.
Crosses: IRAS (Boulanger \& Perault 1988);
squares: COBE-FIRAS (Finkbeiner et al.\ 1999);
diamonds: COBE-DIRBE (Arendt et al.\ 1998);
heavy curve for 3-4.5$\micron$ and $5-11.5\micron$:
IRTS (Onaka et al.\ 1996, Tanaka et al.\ 1996).
The total power $\sim5.1\times10^{-24}\erg\s^{-1}/{\rm H}$ is
estimated from the interpolated broken line.}
\label{fig:IR+FIR_emission}
\end{figure}

Figure \ref{fig:IR+FIR_emission} shows the average emission spectrum
of interstellar dust, based on observations of the FIR emission at
high galactic latitudes, plus 
observations of a section of the galactic plane where the surface
brightness is high enough to permit spectroscopy by the IRTS
satellite (Onaka et al.\ 1996; Tanaka et al.\ 1996).

The similarity of the 5--15$\micron$
spectrum with that of reflection nebulae is evident 
(see Figure \ref{fig:PAH}).
Approximately 21\% of the total power is radiated between 3 and 12$\micron$,
with another $\sim14\%$ between 12 and 50$\micron$.
This emission is from dust grains that are so small that single-photon
heating (see Figure \ref{fig:Thist}) is important.
The remaining $\sim$65\% of the power is radiated in the far-infrared,
with $\lambda I_\lambda$ peaking at $\sim130\micron$.
At far-infrared wavelengths, the grain opacity varies as $\sim\lambda^{-2}$,
and $\lambda I_\lambda \propto \lambda^{-6}/(e^{hc/\lambda kT_d}-1)$
peaks at $\lambda = hc/5.985kT_d = 134\micron (18\K/T_d)$.
The emission spectrum for 18~K dust with opacity $\propto\lambda^{-2}$
shown in Figure \ref{fig:IR+FIR_emission}, provides a good fit to 
the observed spectrum for $\lambda \geq80\micron$, but falls far below
the observed emission at $\lambda \leq 50\micron$.
From Figure \ref{fig:IR+FIR_emission} it is apparent that $\sim60\%$ of
the radiated power appears to originate from grains 
which are sufficiently large
(radii $a\gtsim 100$\AA\ ) so that 
individual photon absorption events do not appreciably raise
the grain temperature.

\section{Dust Grain Size Distribution}

\begin{figure}
\centering
\includegraphics[width=7cm,angle=270]
{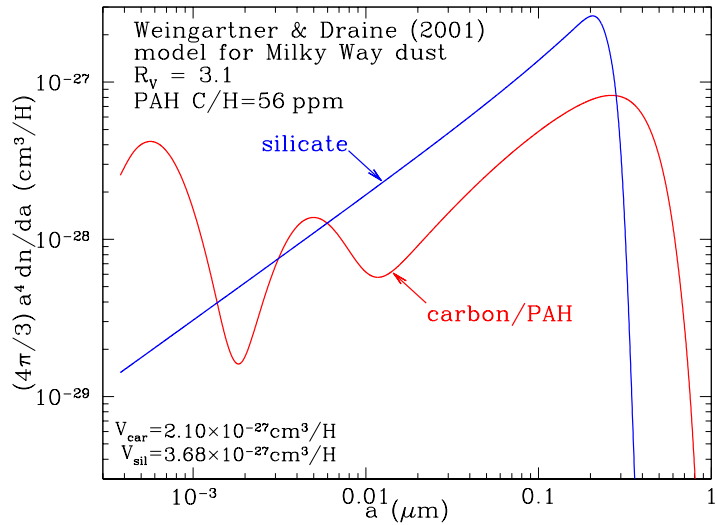}
%{f7.eps}
\caption{Size distributions for dust
	model of Weingartner \& Draine (2001a).
	}
\label{fig:dnda}
\end{figure}

A physical grain model of PAHs, carbonaceous grains, and amorphous silicate
grains has been constructed to reproduce observations of
interstellar extinction and infrared emission 
(Weingartner \& Draine 2001a; Li \& Draine 2001a).
The material dielectric functions and heat capacities, and absorption
cross sections for the PAHs, are consistent with
laboratory data and physics; once the size distribution is specified,
the properties of the grain model (scattering, extinction, infrared
emission) can be calculated.

The grain size distribution for average diffuse clouds ($R_V\approx3.1$ 
in our
region of the Milky Way is shown in Figure \ref{fig:dnda}.
The mass distribution must peak near $\sim0.3\micron$ in order to
reproduce the observed extinction near visual wavelengths.
The peak in the PAH distribution near $.0005\micron$ is required for
single-photon heating to
reproduce the observed 3--12$\micron$ emission, but the
secondary peak near $.005\micron$ in Figure \ref{fig:dnda}
may be an artifact of the fitting
procedure.

Milky Way 
extinction curves with different $R_V$ values, or extinction curves
for the LMC and SMC, can be reproduced by
varying the size distribution.  
The grain model appears to be consistent with the observed scattering
properties of interstellar dust in the optical and ultraviolet
(Draine 2003b)
and at X-ray energies (Draine 2003c).

The grain model also reproduces the observed infrared emission in the
diffuse ISM of the Milky Way (Li \& Draine 2001a), reflection nebulae
(Li \& Draine 2002b),
and the SMC Bar (Li \& Draine 2002c).

\section{Dust Abundances vs. Depletion Patterns}

Jenkins (2003)
finds that observed depletions on many sightlines
suggest two types of grain material:
a ``core'' material which is returned to the gas phase only in very
high velocity shock waves, and a ``mantle'' material which is more easily
stripped.  For low velocity gas, variations in depletion from one
sightline to another are interpreted as due to invariant grain cores
plus varying amounts of mantle
material.

What
do Jenkins' abundances of elements in the cores and mantles suggest
as regards the volumes of cores and mantles?  
Table \ref{tab:stochiometry}
shows one possible mix of minerals which could reproduce the overall
elemental compositions deduced by Jenkins
for sightlines with ``depletion multiplier'' $F_\star=1$,
representative of ``cool disk'' material.  The aims of this exercise are
to estimate (1) how much oxygen is likely to be in the grain material,
and (2) the volumes of core and mantle material which would
be consistent with the depletion patterns found by Jenkins, for comparison
with the grain volume required by physical dust models.
The compositions in Table \ref{tab:stochiometry} are purely illustrative,
and should {\it not} be taken to be realistic.

In Table \ref{tab:stochiometry} the overall composition of the core material
is consistent with Jenkins' results.
The amounts of Mg, Fe, and Si in the mantle require $\sim$57~ppm
O if combined into pyroxene --
consistent with the O/H = $81_{-58}^{+54}$~ppm which Jenkins
assigns to the mantle.

\begin{table}
  \caption{Jenkins (2003) Grain Composition: One Illustrative Possibility}
{\footnotesize
    \begin{tabular}
	{@{}l
	@{\hspace*{1.0em}}c
	@{\hspace*{1.0em}}c
	@{\hspace*{1.0em}}c
	@{\hspace*{1.0em}}c
	@{\hspace*{1.0em}}c
	@{\hspace*{1.0em}}c
	@{\hspace*{1.0em}}c
	@{\hspace*{1.0em}}c
	@{\hspace*{1.0em}}c
	@{\hspace*{1.0em}}c@{}
	}
     \hline \hline
\hspace*{2em}Material		
&C$~^a$	&O$~^a$	&Mg$~^a$ &Si$~^a$ &Al$~^a$ &Ca$~^a$ &Fe$~^a$ &Ni$~^a$ &$\rho~^b$ &$V~^c$\\
\hline\hline
{\bf Grain Cores}\\
C,PAH,HAC,...	
&71	&-	&-	&-	&-	&-	&-	&-	&2.2	&6.5\\
MgFeSiO$_4$ olivine	
&-	&52	&13	&13	&-	&-	&13	&-	&3.8	&9.8\\
CaMgSiO$_4$ monticellite
&-	&8	&2	&2	&-	&2	&-	&-	&3.2	&1.6 \\
Fe$_2$O$_3$ hematite
&-	&18	&-	&-	&-	&-	&12	&-	&5.3	&3.0\\
Al$_2$O$_3$ corundum
&-	&4.5	&-	&-	&3	&-	&-	&-	&4.02	&0.6 \\
%NiO bunsenite
%&-	&1.6	&-	&-	&-	&-	&-	&1.6	&6.67	&0.3\\
Ni$_2$O$_3$ dinickel trioxide
&-	&2.4	&-	&-	&-	&-	&-	&1.6	&4.84	&0.5\\
\hline
Illustrative Core Total	
&71	&85	&15	&15	&3	&2	&25	&1.6	&3.5	&22.1\\
Observed Core Total$~^d$
&71$_{-71}^{+61}$
&53$_{-53}^{+49}$
&15%$_{-1}^{+1}$
&14%$_{-2}^{+1}$
&3.0
&2.2
&25%$_{-0.2}^{+0.2}$
&1.6%$_{-0.1}^{+0.1}$
\\
\hline\hline
{\bf Grain Mantles}\\
C,PAH,HAC,...	
&35	&-	&-	&-	&-	&-	&-	&-	&2.2	&3.2\\
Mg$_{0.9}$Fe$_{0.1}$SiO$_3$ pyroxene
&-	&57	&17	&19	&-	&-	&2	&-	&3.3	&9.9\\
\hline
Illustrative Mantle Total	
&35	&57	&17	&19	&-	&-	&2	&-	&3.5	&13.1\\
\hline\hline
{\bf Cores + Mantles}\\
C,PAH,HAC,...
&106	&-	&-	&-	&-	&-	&-	&-	&2.2	&9.7$~^e$\\
silicates	
&-	&117	&32	&34	&-	&2	&15	&-	&3.5	&21.4$~^e$\\
other	
&-	&24	&-	&-	&3	&-	&12	&1.6	&5.2	&4.0\\
\hline
Illustrative Core + Mantle Total
&106	&142	&32	&34	&3	&2	&27	&1.6	&3.5	&35.2$~^e$\\
Observed Core + Mantle Total$~^d$
&106$_{-20}^{+16}$
&134$_{-23}^{+22}$
&32%32.0$_{-0.2}^{+0.2}$
&33%32.9$_{-0.5}^{+0.4}$
&3.0
&2.2
&28
&1.8\\
\hline\hline
\end{tabular}
}
\label{tab:stochiometry}

{\footnotesize
$^a$ Atomic abundance (ppm) per total H.\\
$^b$ Solid density (${\rm g} \cm^{-3}$).\\
$^c$ Grain volume per total H ($10^{-28}\cm^3$).\\
$^d$ From Jenkins (2003).  Quoted uncertainties do not include uncertainties
        in assumed total abundances.
$^e$ Models that reproduce the observed interstellar extinction per H
	require a greater volume of grain material than provided by
	the depletions found by Jenkins (2003).
	A model based on carbonaceous grains plus silicate grains 
	(see Draine 2003a)
	has
	$V=37\times10^{-28}\cm^3/$H for silicate grains and
	$V=21\times10^{-28}\cm^3$/H for carbonaceous grains (see text).
}
\end{table}

Studies of total depletions infer the amount of
grain material based on an assumed value for the total abundance
of each element; Jenkins
adopted current estimates for solar abundances.  It is important to
remember that estimates of solar abundances have varied considerably over
time; the review by Anders \& Grevesse (1989) had
C/H= $363\pm35$~ppm and O/H = $851\pm72$~ppm, whereas the most recent
redeterminations find
C/H=$246\pm23$~ppm (Allende Prieto et al.\ 2001) and O/H = $490\pm47$~ppm
(Allende Prieto et al.\ 2002) -- C/H has gone down by a factor 1.5,
and O/H by a factor 1.7!
Tomorrow's ``solar abundance'' values may differ from today's.
Furthermore, interstellar abundances may not be equal to solar, or even
to the abundances in young stars, as discussed in \S\ref{sec:transport}
below.

The total grain volume indicated by Jenkins' abundance estimates is
$35\times10^{-28}\cm^3/$H, only $\sim$60\% of the total grain volume for
the physical dust model of Weingartner \& Draine (2001a).
The dust modeling
has approximated the dust as solid spheres, and it is expected that
nonspherical porous grains would allow the extinction to be
reproduced with a slightly reduced total solid volume, but it isn't
clear that this would reduce the
required solid volume by 40\%.

One is compelled to consider seriously the possibility that interstellar
abundances of the depleted elements -- especially C -- 
may exceed solar abundances.
This will be discussed further in \S\ref{sec:transport}.

\section{Time Scale for Depletion}

Lifetimes of grains in H~I clouds against destruction
in supernova blast waves are only $\sim3\times10^8\yr$
(e.g, Draine \& Salpeter 1979,
Jones et al 1994).
How then can we understand the very small
gas-phase abundances routinely found for elements like Si?
The kinetics of depletion have been studied by Weingartner \& Draine (1999),
who show that the population of very small grains 
is capable of accreting metal ions rapidly enough to achieve the required
depletions, provided that interstellar material is rapidly cycled 
between dense clouds and diffuse clouds.

\section{Local Variations in D/H: Does Dust Play a Role?}

The evidence concerning primordial abundances of D, He, and Li appear to
be consistent with D/H$\approx28\pm4$~ppm produced 
by nucleosynthesis
in the early universe (O'Meara et al.\ 2001, Kirkman et al.\ 2003).
Observations of D/H in the Milky Way have been reviewed by Linsky (2003).
The local ISM has a weighted
mean D/H$=15.2\pm0.8$~ppm (1-$\sigma$ uncertainties)
within $\sim$180 pc of the Sun
(Moos et al.\ 2002).
This value could be consistent with ``astration'' if $\sim$50\% of 
the H atoms now in the interstellar medium have previously been in
stars which burnt D to $^3$He.
Observations appear to find spatial variations in
the D/H ratio in the interstellar medium within $\sim500$~pc, with values
ranging from 
$7.4_{-1.3}^{+1.9}$~ppm toward $\delta$ Orionis 
(Jenkins et al.\ 1999) to
$21.8_{-3.1}^{+3.6}$~ppm toward $\gamma^2$ Vel
(Sonneborn et al.\ 2000).
On longer sightlines in the Galactic disk, Hoopes et al.\ (2003) find
D/H $=7.8_{-1.3}^{+2.6}$~ppm toward HD 191877 ($d=2200\pm550$~pc)
and $8.5_{-1.2}^{+1.7}$~ppm toward HD 195965 ($d=800\pm200$~pc).
These variations in D/H are usually interpreted as indicating variations
in ``astration'', with as much as
$\sim75\%$ of the D on the sightline to $\delta$ Ori having been ``burnt'',
vs.\ only $\sim$20\% of the D toward $\gamma^2$Vel.
Such large variations in astration between regions situated just a few
hundred pc apart would be surprising, since the ISM
appears to be sufficiently well-mixed that large local variations in 
the abundances of elements like N or O are not seen outside of
recognizable stellar ejecta such as planetary nebulae or supernova remnants.

Jura (1982) pointed out that interstellar grains could conceivably
sequester a significant amount of D.
Could the missing deuterium conceivably be in dust grains?

Let us suppose that dust grains contain 200~ppm C relative to total H
(as in the dust model of Weingartner \& Draine 2001)
with $\sim$60~ppm 
in PAHs containing $N_C \ltsim 10^4$ atoms.

The solid carbon will be hydrogenated to some degree.
The most highly pericondensed PAH molecules
(coronene ${\rm C}_{24}{\rm H}_{12}$, 
circumcoronene ${\rm C}_{54}{\rm H}_{18}$, dicircumcoronene
${\rm C}_{96}{\rm H}_{24}$) have H/C = $\sqrt{6/N_{\rm C}}$,
where $N_{\rm C}$ is the number of C atoms; other PAHs have higher H/C ratios
for a given $N_{\rm C}$.
Let us suppose that the overall carbon grain material -- including small
PAHs and larger carbonaceous grains -- has H/C=0.25.

The carbonaceous grain population would then contain $\sim$50~ppm of hydrogen.
If $\sim$20\% of the hydrogen in the carbonaceous
grains was deterium, the deuterium in the grains would then be
(D)$_{grain}$/(H)$_{total} \approx 10$~ppm.
If the total D/H = 20~ppm, this would reduce the gas phase D/H to 10~ppm,
comparable to the value observed toward $\delta$Ori.

Is it conceivable the D/H ratio in dust grains could be
$\sim10^4$ times higher than the overall D/H ratio?
Some interplanetary dust particles have D/H as high as .0017 
(Messenger \& Walker 1997, Keller et al.\ 2000), although this factor
$\sim$85 enrichment (relative to D/H=20~ppm) is still two orders of magnitude
less than what is required to significantly affect the gas phase D/H value.
Extreme D enrichments are seen in some 
interstellar molecules -- D$_2$CO/H$_2$CO ratios in the range of .01 - 0.1
are seen (Ceccarelli et al 2001; Bacmann et al 2003), 
and attributed to
chemistry on cold grain surfaces in dense clouds.

Could such extreme enrichments occur in the diffuse interstellar medium?
The thermodynamics is favorable.
The H or D would be bound to the carbon via
a C-H bond.  The C-H bond -- with a bond strength $\sim3.5\eV$ --
has a stretching mode at $\lambda_{\rm CH}=3.3\micron$,
while the C-D bond, with a larger reduced mass, has its stretching mode
at $\lambda_{\rm CD}\approx\sqrt{2}\lambda_{\rm CH}\approx 4.67\micron$.
Because of the difference in zero-point energy, the difference in
binding energies is
%\begin{eqnarray}
%\Delta E_{\rm CD-CH} &=& 
%\frac{hc}{2}\sum_{j=1}^3(\lambda_{{\rm CH},j}^{-1}-\lambda_{{\rm CD},j}^{-1})
%\\
%&\approx& \frac{hc}{2}\left(1-\frac{1}{\sqrt{2}}\right) 
%\sum_{j=1}^3\lambda_{{\rm CH},j}^{-1}= .092\eV
%\end{eqnarray}
\begin{equation}
\Delta E_{\rm CD-CH} =
\frac{hc}{2}\sum_{j=1}^3(\lambda_{{\rm CH},j}^{-1}-\lambda_{{\rm CD},j}^{-1})
\approx \frac{hc}{2}\left(1-\frac{1}{\sqrt{2}}\right) 
\sum_{j=1}^3\lambda_{{\rm CH},j}^{-1}= .092\eV~,
\end{equation}
where the sum is over the stretching, in-plane bending, and out-of-plane
bending modes, with $\lambda_{\rm CH}=3.3$, 8.6, and 11.3$\micron$.
This exceeds the
difference $\Delta E_{\rm HD-H_2}=.035\eV$ in
binding energy between HD and H$_2$.
It is therefore energetically
favored for impinging D atoms to displace bound H atoms via
reactions of the form (Bauschlicher 1998)
\begin{eqnarray}
{\rm C}_l{\rm D}_m{\rm H}_n^+ + {\rm H} 
&\rightarrow& 
{\rm C}_l{\rm D}_m{\rm H}_{n+1}^+
\\
{\rm C}_l{\rm D}_m{\rm H}_{n+1}^+ + {\rm D} 
&\rightarrow& 
{\rm C}_l{\rm D}_{m+1}{\rm H}_{n-1}^+ + {\rm H}_2
~~~~~~{\rm branching~fraction~}f_1
\label{eq:reaction1}
\\
{\rm versus}\hspace*{6em}&\rightarrow&
{\rm C}_l{\rm D}_{m}{\rm H}_{n}^+ + {\rm HD}
~~~~~~~~~~{\rm branching~fraction~}f_2 ~~.
\label{eq:reaction2}
\end{eqnarray}
The branching ratio
$f_1/f_2\approx\exp[(\Delta E_{\rm CD-CH}-\Delta E_{\rm HD-H_2})/kT_{\rm d}] > 10^4$ 
if $T_{\rm d}\ltsim70\K$.
If no other reactions affect the grain hydrogenation, then
the grains would gradually become D-enriched.

However, the interstellar medium is far from LTE --
the hydrogen is atomic (rather than molecular) 
and, indeed, partially ionized because of the
presence of ultraviolet photons, X-rays, and cosmic rays. 
It is not yet clear whether the mixture of non-LTE reactions will allow
the grains to become deuterated to a level approaching D/H $\approx$ 1/4,
but it seems possible that this may occur.
Deuterated PAHs would radiate in the C-D stretching and bending modes
at $\sim4.67$, 12.2, and 16.0$\micron$.  The 12.2$\micron$ emission will
be confused with C-H out-of-plane bending emission (see Figure \ref{fig:PAH}),
but the other two modes should be searched for.

Even if extreme D-enrichment of carbonaceous grains is possible,
it will take time to develop.  Meanwhile, the gas in which the grain
is found may undergo a high velocity shock, with grain destruction by
a combination of sputtering and ion field emission
in the high temperature postshock gas.
D incorporated into dust grains would be released and returned to the
gas phase if those dust grains are destroyed.  PAHs, in particular,
would be expected to be easily sputtered in shock-heated gas;
destruction by ion field emission would be expected to be even more
rapid.
Thus if D is depleted into dust grains, we would expect to see
the gas-phase D/H to be larger in recently-shocked regions.
This could explain the large D/H value observed by
Sonneborn et al.\ (2000) toward $\gamma^2$Vel.

It should be noted that significant depletion of D from the gas 
can only occur if there is sufficient carbonaceous grain material
to retain the D.  This can occur if gas-phase abundances are
approximately solar (with $\sim$200~ppm C in dust)
but would not be possible for abundances significantly below solar -- e.g.,
the abundances in the LMC and SMC.  
The factor-of-two variations in D/H seen in the local
interstellar medium would {\it not} be possible in gas with
metallicities characteristic of the LMC, SMC, or high-velocity clouds
such as ``complex C'' (see Jenkins 2003).

\section{Transport of Elements in Dust Grains
	\label{sec:transport}}

Elements like Mg, Si, Al, Ca, Ti, Fe, Ni are concentrated in interstellar
dust grains.  Since dust grains can move through the gas,
transport in dust grains could produce
local variations in elemental abundances.

\subsection{Anisotropic Starlight}

Starlight is typically anisotropic, as can be confirmed by viewing
the night sky on a clear night.  
The anisotropy is a function of wavelength,
with larger anisotropies at shorter wavelengths 
because UV is (1) more strongly attenuated by dust grains and 
(2) originates in a smaller number of
short-lived stars that are clustered.
At the location of the
Sun, Weingartner \& Draine (2001b) found starlight anistropies ranging from
3\% at 5500\AA\ to 21\% at 1565\AA.

The dust grains are charged, and fairly well-coupled to the magnetic
field lines.
Let $\theta_B$ be the angle between the starlight anisotropy direction
and the magnetic field.
The drift tends to be approximately parallel to the magnetic field
direction, with a magnitude $v_d\approx v_{d0}\cos\theta_B$; for a starlight
anisotropy of 10\%, 
$v_{d0}\approx 0.5\kms$ in the ``warm neutral medium'', and
$0.03\kms$ in the ``cold neutral medium'' (Weingartner \& Draine 2001b).

If the radiation anisotropy is stable for $\sim10^7\yr$, a dust grain
in the cold neutral medium
would be driven 0.3$\cos\theta_B$ pc 
from the gas element in which it 
was originally located.

If the magnetic field is nonuniform, spatial variations in the drift
velocity can then lead to variations in the dust/gas ratio.
Indeed, if there are ``valleys'' in the magnetic field (with respect
to the direction of radiation anisotropy'', the dust grains would
tend to be concentrated there.  Radiation pressure acting on the
concentrated dust grains would in fact cause such perturbations in
the magnetic field to be unstable.

Dust impact detectors on the Ulysses and Galileo spacecraft measure 
the flux and mass distribution of
interstellar grains entering the solar system, finding
a much higher flux of very large ($a\gtsim 0.5\micron$ dust grains
than would be expected for the average interstellar grain size
distribution (Frisch et al.\ 1999).
The inferred local dust size distribution is very difficult to
reconcile with the average interstellar grain population
(Weingartner \& Draine 2001a), but we must keep in mind that
the dust grains entering the solar system over a time scale of
a few years are sampling a
tiny region of the interstellar medium of order a few AU in size.
It is possible that the solar system just happens to be passing
through a region in the local interstellar cloud which has been
enhanced in the abundance of large grains due to radiation-driven
grain drift.

\subsection{Star-Forming Regions}

Various processes could alter
the gas-to-dust ratio in the material forming a star:
\begin{enumerate}
\item 
Motion of dust through gas can result from gravitational sedimentation
of dust grains in star-forming clouds (Flannery \& Krook 1978).
This could lead to enhanced abundances in stars of those elements
which are depleted into dust in star-forming clouds.
\item
Star formation is accompanied by
ambipolar diffusion of magnetic field out of the contracting gas cloud.
Charged dust grains, while not perfectly coupled to the magnetic field,
will tend to drift outward, resulting in reduction of the abundances
of the depleted elements in the star-forming core
(Ciolek \& Mouschovias 1996).
\item
Gravitational sedimentation in an accretion disk can concentrate dust
at the midplane.
Gammie (1996) has suggested ``layered accretion'', in which 
viscous stresses are effective along the surface layers of the disk,
but the midplane is a quiescent ``dead zone'' as far as the magnetorotational
instability is concerned.  This could suppress stellar abundances of
depleted elements.
\item
In thick disks, small bodies orbit more rapidly than the gas, with gas
drag leading to radial infall of these bodies.  This could enhance
stellar abundances of depleted species.
\item
Before accretion terminates, a massive star may attain
a pre-main-sequence luminosity high enough for radiation pressure to drive
a drift of dust grains away from it.  This would suppress stellar
abundances of depleted elements.
\end{enumerate}
Given these competing mechanisms,
the net effect could be of either sign,
but, as pointed out by Snow (2000), it would not be suprising if
stars had abundances which differed from the overall abundances
of the interstellar medium out of which they formed, and it would
also not be surprising if the resulting abundances depended on stellar mass.
In this connection it is interesting to note that the study by
Sofia \& Meyer (2001) finds that abundances of Mg and Si in B stars
appear
to be significantly below the abundances of these elements in young
F and G disk stars,
or in the Sun; they conclude that the abundances in B stars do not provide
a good representation of interstellar abundances.

\section{Ion Recombination on Dust Grains}

As discussed by Jenkins (2003), determinations of elemental abundances
from interstellar absorption line observations may require
estimation of the fraction in unobserved ionization states,
such as Na~II, K~II, or Ca~III.
Such estimates require knowledge of the rates for ionization and
recombination processes.
Neutral or negatively-charged dust grains provide a pathway for
ion neutralization that for some cases can be faster than ordinary radiative
recombination with free electrons
(Weingartner \& Draine 2002).

\section{Summary}

This symposium has been concerned with the origin and evolution of the
elements.  The main points of the present paper are as follows:
\begin{enumerate}
\item The chemical composition of interstellar dust remains uncertain.
PAH molecules are present,
and a substantial fraction of the grain mass most likely consists
of amorphous silicate, but precise compositional information still eludes
us.
\item The grain size distribution is strongly constrained by observations
of extinction, scaterring, and infrared emission, and
extends from grains containing just
tens of atoms to grains with radii $a\gtsim0.3\micron$.
\item Dust grains will drift through interstellar gas, and this transport
process could produce local variations in the dust/gas ratio.
\item The dust mass required to account for interstellar extinction using
homogeneous spherical grains exceeds that 
inferred from depletion studies (Jenkins 2003)
by a factor $\sim1.5$.
\item Dust grains could possibly deplete a significant fraction of
interstellar D.  This mechanism should be considered as a possible
explanation for observed variations in the interstellar D/H ratio.
\end{enumerate}
\section{Acknowledgements}

I thank Ed Jenkins and Todd Tripp for many valuable discussions, and 
Robert Lupton for availability of the SM software package.
This research was supported in part by NSF grant AST-9988126.


\begin{thereferences}{}

\def\aa{{A\&A}}
\def\aas{{A\&AS}}
\def\aj{{AJ}}
\def\annrev{{ARA\&A}}
\def\apj{{ApJ}}
\def\apjs{{ApJS}}
\def\baas{{BAAS}}
\def\jgr{{J.~Geophys.~Res.~}}
\def\gca{{Geochim.~Cosmochim.~Acta}}
\def\lickobsbull{{Lick Obs.\ Bull.~}}
\def\mnras{{MNRAS}}
\def\nat{{Nature}}
\def\pasp{{PASP}}
\def\ssr{{Space Sci.\ Rev.}}

\bibitem{AP01}
	% The Forbidden Abundance of Oxygen in the Sun
	Allende Prieto, C., Lambert, D.L., \& Asplund, M. 2001
	\apj, 556, L63

\bibitem{AP02}
	% A Reappraisal of the Solar Photospheric C/O Ratio
	Allende Prieto, C., Lambert, D.L., \& Asplund, M. 2002,
	\apj, 573, L137

\bibitem{AG89}
        Anders, E.A., \& Grevesse, N. 1989,
	Geochim. Cosmochim. Acta, 53, 197

\bibitem{Ar98}
        Arendt, R.G., Odegard, N, Weiland, J.L., Sodroski, T.J.,
	Hauser, M.G., 
	%Dwek, E., Kelsall, T., 
	et al.\ 1998,
	\apj, 508, 74

\bibitem{Ar00}
	Artymowicz, P. 2000,
	\ssr, 92, 69

\bibitem{BLC03}
	Bacmann, A., Lefloch, B., Ceccarelli, C., Steinacker, J.,
	Castets, A., \& Loinard, L. 2003,
	\apj, 585, L55

\bibitem{Ba98}
	% Reaction of PAH+ with H atoms: Astrophysical Implications
	Bauschlicher, C.W. 1998,
	\apj, 509, L125

\bibitem{BP88}
        Boulanger, F., \& Perault, M. 1988,
	\apj, 330, 964

\bibitem{BMK01}
	Bouwman, J., Meeus, G., de Koter, A., Hony, S., Dominik, C.,
	\& Waters, L.B.F.M. 2001,
	\aa, 375, 950

\bibitem{BA02}
	Bowey, J.E., \& Adamson, A.J. 2002,
	\mnras, 334, 94

\bibitem{BAW98}
	Bowey, J.E., Adamson, A.J, \& Whittet, D.C.B. 1998,
	\mnras, 298, 131

\bibitem{CLC01}
	Ceccarelli, C., Loinard, L., Castets, A., Tielens, A.G.G.M.,
	Caux, E., Lefloch, B., Vastel, C. 2001,
	\aa 373, 998

\bibitem{CCM89}
        Cardelli, J.A., Clayton, G.C., \& Mathis, J.S. 1989,
	\apj, 345, 245

\bibitem{CLA96}
	Cesarsky, D., Lequeux, J., Abergel, A., Perault, M., Palazzi, E.,
	et al.\ 1996,
	\aa, 315, L305

\bibitem{CM96}
	% the effect of ambipolar diffusion on the dust-to-gas ratio
	% in protostellar cores
	Ciolek, G.E., \& Mouschovias, T. C. 1996,
	ApJ, 468, 749

\bibitem{CN03}
	Clayton, D.D., \& Nittler, L.R. 2003, 
	in {\it Carnegie Observatories Astrophysics Series, Vol.\ 4:
	Origin and Evolution of the Elements},
	ed. A. McWilliam \& M. Rauch
	(Cambridge: Cambridge Univ. Press), 000

\bibitem{DPS99}
	Darbon, S., Perrin, J.M., \& Sivan, J.P. 1999,
	\aa, 348, 990

\bibitem{DZP00}
	Darbon, S., Zavagno, A., Perrin, J.M., Savine, C., Ducci, V., 
	\& Sivan, J.P. 2000,
	\aa, 364, 723

\bibitem{DLO86}
	d'Hendecourt, L.B., L\'eger, A., Olofson, G., \& Schmidt, W. 1986,
	\aa, 170, 91

\bibitem{Dr89}
	Draine, B.T. 1989, in
	{\it Interstellar Dust, Proc. of IAU Symp. 135},
	ed. L.J. Allamandola \& A.G.G.M. Tielens
	(Dordrecht: Kluwer), 313

\bibitem{Dr90}
	Draine, B.T. 1990,
	in {\it Evolution of the Interstellar Medium},
	ed. L. Blitz
	(San Francisco: Astr. Soc. Pac.),
	193

\bibitem{Dr03a}
	Draine, B.T. 2003a,
	\annrev, 41, 241

\bibitem{Dr03b}
	Draine, B.T. 2003b, 
	% Scattering by Interstellar Dust Grains. I. Optical and Ultraviolet
	\apj, 598, 1017

\bibitem{Dr03c}
	Draine, B.T. 2003c, 
	% Scattering by Interstellar Dust Grains. II. X-Rays
	\apj, 598, 1026

\bibitem{Dr03d}
	Draine, B.T. 2004,
	% Astrophysics of Dust in Cold Clouds
	in {\it The Cold Universe, Saas-Fee Advanced Course 32},
	ed. D. Pfenniger
	(Berlin: Springer-Verlag),
	p. 213
	(astro-ph/0304489)

\bibitem{DL01}
        Draine, B.T., \& Li, A. 2001,
	\apj, 551, 809

\bibitem{DS79}
	Draine, B.T., \& Salpeter, E.E. 1979,
	\apj, 231, 438

%\bibitem{Dul85}
%	Duley, W.W. 1985,
%	\mnras, 215, 259

\bibitem{DSS98}
	Duley, W.W., Scott, A.D., Seahra, S., \& Dadswell, G. 1998,
	\apj, 571, L117

\bibitem{FM90}
	% An analysis of the shapes of ultraviolet extinction curves. 
	% III - an atlas of ultraviolet extinction curves
	Fitzpatrick, E.L., \& Massa, D. 1990,
	\apjs, 72, 163
	
\bibitem{FDS99}
        Finkbeiner, D.P., Davis, M., \& Schlegel, D.J. 1999,
	\apj, 524, 867

\bibitem{Fi99}
        Fitzpatrick, E.L. 1999,
	\pasp, 111, 63

\bibitem{FK78}
	% The sedimentation of grains in interstellar clouds
	Flannery, B.P, \& Krook, M. 1978,
	\apj, 223, 447

\bibitem{FE94}
	Foing, B.H., \& Ehrenfreund, P. 1994,
	\nat, 369, 296

\bibitem{FWC98}
        Forrey, R.C., Woo, J.W., \& Cho, K. 1998,
	\apj, 505, 236

\bibitem{Fea99}
	Frisch, P.C. et al.\ 1999,
	\apj, 525, 492

\bibitem{FW90}
	Furton, D.G., \& Witt, A.N. 1990,
	\apj, 364, L45

\bibitem{GKM00}
	% On the identification of the C60+ interstellar features
	Galazutdinov, G.A., Krelowski, J., Musaev, F.A., Ehrenfreund, P.,
	\& Foing, B.H. 2000,
	\mnras, 317, 750

\bibitem{Ga96}
	% Layered accretion in T Tau disks
	Gammie, C.F. 1996,
	\apj, 457, 355

\bibitem{GWF98}
	Gordon, K.D., Witt, A.N., \& Friedmann, B.C. 1998,
	\apj, 498, 522

\bibitem{Ha99}
	Hanner, M. 1999,
	\ssr, 90, 99

\bibitem{He22}
	% The Spectra of Certain Class B Stars in the Regions 5630A-6680A
	% and 3280A-3380A
	Heger, M.L.
	1922,
	\lickobsbull 10, 146

\bibitem{HSH03}
	% Deuterium toward Two Milky Way Disk Stars: Probing Extended Sight 
	% Lines with the Far Ultraviolet Spectroscopic Explorer
	Hoopes, C.G., Sembach, K.R., H\'ebrard, G., Moos, H.W., Knauth, D.C.
	2003,
	ApJ, 586, 1094

\bibitem{Je03}
	Jenkins, E.B. 2003,
	in {\it Carnegie Observatories Astrophysics Series, Vol.\ 4:
	Origin and Evolution of the Elements},
	ed. A. McWilliam \& M. Rauch
	(Cambridge: Cambridge Univ. Press), 000

\bibitem{JTW99}
	% Spatial Variability in the Ratio of Interstellar Atomic Deuterium 
	% to Hydrogen. I. Observations toward delta Orionis by the 
	% Interstellar Medium Absorption Profile Spectrograph
	Jenkins, E.B., Tripp, T.M., Wozniak, P.R., Sofia, U.J.,
	\& Sonneborn, G. 1999, \apj, 520, 182

\bibitem{JD94}
	% A survey of diffuse interstellar bands (3800-8680A)
	Jenniskens, P., \& Desert, F.X.
	1994,
	\aas, 106, 39

\bibitem{JMP97}
	% Diffuse interstellar bands near 9600A: not due to C60+ yet
	Jenniskens, P., Mulas, G., Porceddu, I, \& Benvenuti, P. 1997,
	\aa, 327, 337

\bibitem{JTH94}
	Jones, A.P., Tielens, A.G.G.M., Hollenbach, D.J., \& McKee, C.F. 1994,
	\apj, 433, 797

\bibitem{Ju82}
	Jura, M. 1982,
	in Advances in Ultraviolet Astronomy,
	ed. Y. Kondo,
	(NASA CP-2238), 54

\bibitem{KMB00}
	% Analysis of a deuterium-rich IDP and implicatoins for presolar
	% material in IDPs
	Keller, L.P., Messenger, S., \& Bradley, J.P. 2000,
	\jgr, 105, 10397

\bibitem{KHF98}
	%
	Kerr, T.H., Hibbins, R.E., Fossey, S.J., Miles, J.R., 
	\& Sarre, P.J.
	1998,
	\apj, 495, 941

\bibitem{KHM96}
	% Molecular rotational contour fitting of ultra-high-resolution 
	% profiles of diffuse interstellar bands
	Kerr, T.H., Hibbins, R.E., Miles, J.R., Fossey, S.J., 
	Sommerville, W.B., \& Sarre, P.J.
	1996,
	\mnras, 283, 1104

\bibitem{KTS03}
	% The cosmological baryon density fromt the D/H ratio towards
	% QSO absorption systems: D/H toward Q1243+3047
	Kirkman, D., Tytler, D., Suzuki, N., O'Meara, J.M., \& Lubin, D. 2003,
	\apjs, submitted (astro-ph/0302006)

\bibitem{LEG98}
	Ledoux, G., Ehbrecht, M., Guillois, O., Huisken, F., Kohn, B., 
	et al.\ 1998,
	\aa, 333, L39

\bibitem{LD01a}
	% On Ultrasmall silicate grains in the diffuse ISM
	Li, A., \& Draine, B.T. 2001a,
	\apj, 550, L213

\bibitem{LD01a}
	% IR emission from interstellar dust II.  the diffuse ism
	Li, A., \& Draine, B.T. 2001b,
	\apj, 554, 778

\bibitem{LD02a}
	% Are silicon nanoparticles an interstellar dust component?
	Li, A., \& Draine, B.T. 2002a,
	\apj, 564, 803

\bibitem{LD02b}
	% Do the IR emission features need uv excitation? The PAH model
	% in UV-poor reflection nebulae
	Li, A., \& Draine, B.T. 2002b,
	\apj, 569, 232

\bibitem{LD02c}
	% IR Emission from Interstellar Dust: III. The SMC
	Li, A., \& Draine, B.T. 2002c,
	\apj, 576, 762

\bibitem{Li03}
	Linsky, J.L. 2003,
	in {\it Carnegie Observatories Astrophysics Series, Vol.\ 4:
	Origin and Evolution of the Elements},
	ed. A. McWilliam \& M. Rauch
	(Cambridge: Cambridge Univ. Press), 000
		
\bibitem{MFB80}
	McCarthy, J.F., Forrest, W.J., Briotta, D.A., \& Houck, J.R. 1980,
	\apj, 242, 965

\bibitem{Me34}
	% Unidentified Interstellar Lines
	Merrill, P.W.,
	1934.
	\pasp, 46, 206

\bibitem{MW97}
	% Evidence for molecular cloud material in meteorites and
	% interplanetary dust,
	Messenger, S., \& Walker, R.M. 1997,
	in {\it Astrophysical Implications of the Laboratory Study of
	Presolar Materials}, ed. T.J. Bernatowitcz \& E.K. Zinner,
	{\it AIP Conf. Proc.}, 402, 545.

\bibitem{MSV02}
	% Abundances of Deuterium, Nitrogen, and Oxygen in the Local 
	% Interstellar Medium: Overview of First Results
	% from the FUSE Mission
	Moos H.W., et al.\ 2002,
	\apjs, 140, 3

\bibitem{MSV99}
	% Upper limits on C60 and C60+ features in the ISO-SWS
	% spectrum of the reflection nebula NGC 7023
	Moutou, C., Sellgren, K., Verstraete, L, \& L\'eger, A. 1999,
	\aa, 347, 949

\bibitem{OMT01}
	O'Meara, J.M., Tytler, D., Kirkman, D., Suzuki, N., Prochaska, J.X.,
	Lubin, D., \& Wolfe, A.M. 2001,
	\apj, 552, 718

\bibitem{OYT96}
	% Detection of the Mid-Infrared Unidentified Bands in the 
	% Diffuse Galactic Emission by IRTS
	Onaka, T., Yamamura, I., Tanabe, T., Roellig, T.L., \& Yuen, L.
	1996.
	\pasj, 48, L59

\bibitem{PBM01}
	% Interstellar X-ray Absorption Spectroscopy of O, Ne, and Fe with
	% the Changra LETGS Spectrum of X0614+091
	Paerels, F., Brinkman, A.C., van der Meer, R.L.J., Kaastra, J.S.,
	Kuulkers, E.,
	et al.\ 
	%, Boggende AJF den, Predehl P, Drake JJ, Kahn SM, 
	% Savin DW, McLaughlin BM.
	2001,
	\apj, 546, 338

\bibitem{RA84}
	%
	Roche, P.F., \& Aitken, D.K. 1984,
	\mnras, 208, 481

\bibitem{SAT91}
	Sandford, S.A., Allamandola, L.J., Tielens, A.G.G.M., Sellgren, K.,
	Tapia, M., \& Pendleton, Y. 1991,
	\apj, 371, 607

\bibitem{SCC02}
	% the first high-resolution X-ray spectrum of Cyg X-1: Soft-Xray
	% ionization and absorption
	Schulz, N.S., Cui, W., Canizares, C.R., Marshall, H.L., 
	Lee, J.C., et al.\ 
	% Miller JM, Lewin WHG.
	2002,
	\apj, 556, 1141

\bibitem{SP93}
	Sivan, J.P., \& Perrin, J.M. 1993,
	\apj, 404, 258

\bibitem{SW02}
	Smith, T.L., \& Witt, A.N. 2002,
	\apj, 565, 304

\bibitem{SWA00}
	Smith, C.H., Wright, C.M., Aitken, D.K., Roche, P.F., 
	\& Hough, J.H. 2000,
	\mnras, 312, 327

\bibitem{Sn00}
	% composition of interstellar gas and dust
	Snow, T.P. 2000,
	\jgr, 105, 10239

\bibitem{STF00}
	% Spatial Variability in the Ratio of Interstellar Atomic Deuterium 
	% to Hydrogen. II. Observations toward gamma2 Velorum and 
	% zeta Puppis by the Interstellar Medium Absorption Profile 
	% Spectrograph
	Sonneborn, G., Tripp, T.M., Ferlet, R., Jenkins, E.B., Sofia, U.J.,
	Vidal-Madjar, A., \& Wozniak, P.R. 2000,
	\apj, 545, 277

\bibitem{SM01}
        Sofia, U.J., \& Meyer, D.M. 2001,
	\apj, 554, L221; 558, L147

\bibitem{SD65}
	Stecher, T.P., \& Donn, B. 1965,
	\apj, 142, 1681

\bibitem{SG98}
	Szomoru, A., \& Guhathakurta, P. 1998,
	\apj, 494, L93

\bibitem{TFM02}
	% O and Ne K absorption edge structures and interstellar abundance
	% towards Cyg X-2
	Takei, Y., Fujimoto, R., Mitsuda, K., \& Onaka, T.
	2002.
	\apj, 581, 307

\bibitem{TMM96}
	% IRTS Observation of the Unidentified 3.3-Micron Band 
	% in the Diffuse Galactic Emission 
	Tanaka, M., Matsumoto, T., Murakami, H., Kawada, M., 
	Noda, M., \& Matsuura, S.
	1996,
	\pasj, 48, L53

\bibitem{WMW00}
	Waelkens, C., Malfait, K., \& Waters, L.B.F.M. 2000,
	in {\it IAU Symp.\ 197, Astrochemistry,} ed.\ Y.C.\ Minh \&
	E.F.\ van Dishoeck (San Francisco: ASP), 435

\bibitem{WD99}
	% Interstellar depletion onto very small dust grains
	Weingartner, J.C., \& Draine, B.T. 1999,
	\apj, 517, 292

\bibitem{WD01a}
	% Dust Grain Size Distributions and Extinction in the
	% Milky Way, LMC, and SMC
	Weingartner, J.C., \& Draine, B.T.
	2001a.
	\apj, 548, 296

\bibitem{WD01b}
	% Forces on dust grains exposed to anisotropic radiation fields
	Weingartner, J.C., \& Draine, B.T. 2001b,
	\apj, 553, 581
	
\bibitem{WD02}
	Weingartner, J.C., \& Draine, B.T. 2002,
	% Electron-Ion Recombination on Grains and Polycyclic Aromatic
	% Hydrocarbons
	\apj, 563, 842

\bibitem{Wh03}
        Whittet, D.C.B. 2003,
	{\it Dust in the Galactic Environment},
	2nd ed.
	(Bristol: IOP)

\bibitem{WB90}
	Witt, A.N., \& Boroson, T.A. 1990,
	\apj, 355, 182

\bibitem{WGF98}
	Witt, A.N., Gordon, K.L., \& Furton, D.G. 1998,
	\apj, 501, L111

\bibitem{WS85}
	Witt, A.N., \& Schild, R.E. 1985,
	\apj, 294, 225

\end{thereferences}
\end{document}